# A Carbon-Ion Superconducting Gantry and a Synchrotron Based on Canted Cosine Theta Magnets.

E. Benedetto[1,3], N. Al Harbi[1], L. Brouwer[2], D. Tommasini[3], S. Prestemon[2], P. Riboni[1] and U. Amaldi[1]

[1] TERA Foundation, Novara, Italy
[2] LBL, Berkeley, USA
[3] CERN, Geneva, Switzerland

E-mail: elena.benedetto@cern.ch

**Abstract**

This article presents the conceptual design of a new compact superconducting gantry and synchrotron for Carbon ion therapy and focuses on the solutions (layout and optics) to make it compact.
The main specificity of this gantry design is to be smaller and lighter with respect to the existing carbon-ion gantries. This is achieved by adopting an innovative mechanical design and by using superconducting magnets of the Canted Cosine Theta type with a small aperture. The optics is optimized to reduce the beam size, it is achromatic and it is independent of the rotation angle (for an incoming round beam).
A preliminary synchrotron layout based on similar superconducting magnets units is presented and the dose delivery specificities are discussed.

## 1 Introduction

The focus of this work is the design of a compact gantry based on superconducting (SC) magnets, which delivers a carbon-ion beam with energies ranging from 100 MeV/u to 430 MeV/u.

The carbon-ion beam rigidity at top energy is $B\rho$ = 6.6 Tm. This corresponds to a bending radius of the order of $\rho$ = 4.5 m for normal conducting (NC) magnets, assuming a maximum field of $B$=1.4 T. Depending on the maximum magnetic field that they can reach, with SC magnets the bending radius can be reduced of at least a factor 2.

The first carbon-ion gantry ever built and still in operation at HIT, Germany, has NC magnets and weighs 600 tons [1]. The second gantry, built at NIRS, Japan, uses SC magnets at 2.9 T. This choice allows reducing its weight to about 300 tons [2]. These two carbon-ion gantries, the only ones in operation so far, have the two fast scanning magnets (SMs) for beam delivery system *upstream* of the last bending magnet, which implies that the aperture of this magnet has to be large enough to allow steering the beam and paint a tumour cross-section of at least 20 cm x 20 cm. To reduce the aperture and the weight of the last dipole, in the gantry here proposed the SMs are *downstream* of the last bending magnet. This is also the approach chosen in the most recent design by NIRS, which is based on 5-tesla bending magnets [3].

Existing synchrotrons for carbon-ion therapy are based on NC magnets and have 55-75 m circumferences [4] [5] with footprints of at least 25 m x 25 m. SC magnets allow the reduction of the synchrotron circumference to less than 30 m and a 8 m x 8 m footprint that is comparable to the one of a NC-magnet proton therapy synchrotron. No SC-magnet synchrotron for medical treatment has been built so far, but several designs exist, in particular the one by NIRS [3].

The layout here presented, similar to the NIRS one, is based on $90^0$ bending magnets (BMs) with two main differences: (i) the magnets are of the Alternating Gradient - Canted Cosine Theta (AG-CCT) type and (ii) a novel gantry mechanics has been introduced. The first choice eases the fabrication of strongly curved magnets and helps in reducing the transverse beam size. As a consequence of the second choice the rotating part of the gantry weighs about 35 tons, which is even smaller than the one of proton gantries [6].

In a CCT magnet a pair of conductor layers are wound and powered such that their axial field components cancel, and their transverse field components sum generating a pure dipole field [7], The extension proposed in [8] foresees two other internal layers (Figure 1) that generate an alternation of focusing and defocusing quadrupolar fields within the magnet and explain the name AG-CCT. The present coil design follows directly the work done at LBNL, in collaboration with PSI [8] [9], to produce a large momentum acceptance $90^0$ magnet for a proton gantry; recently a prototype has been tested [10].

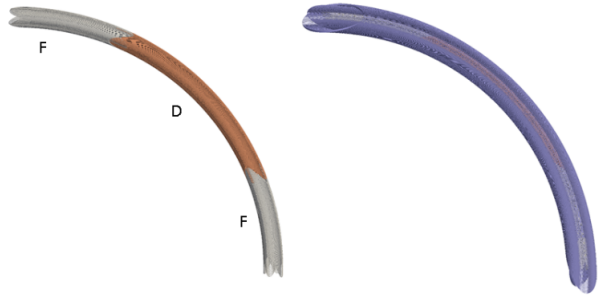

*Figure 1*: *Sketch of the coils. An inner set of AG-CCT windings produces alternating quadrupole fields along the magnet (left). These layers are nested inside an outer set of CCT dipole windings, which produce a constant dipole field along the length.*

## 2 Methods

### 2.1 Superconducting magnets

In the $90^0$ SC magnets of the synchrotron and the gantry the two inner AG-CCT (quadrupole) windings followed and the outer set of CCT (dipole) layers [9] generate the fields specified in Table 1.

*Table 1:* *Specifications for the gantry and the synchrotron magnets*

|  | Gantry | Synchrotron |
|---|---|---|
| Max. bending field | 4.0 T | 3.5 T |
| Bending Radius | 165 cm | 189 cm |
| Magnetic bending angle | 90° | |
| Max. quadrupole gradient | 10 T/m | |
| Radius of vacuum chamber | 20 mm | 30 mm |
| Radius of the inner coil layer | 40 mm | 50 mm |
| Peak field at the coil | 4.4 T | 4.0 T |
| Quadrupole alternation | FDF: 22.5° - 45° - 22.5° | |

A maximum dipole field of 3.5 T for the synchrotron (with a larger aperture) and 4 T for the gantry (with a smaller aperture) have been chosen. The corresponding bending radii are respectively ρ= 1.89 m and ρ= 1.65 m.

The 22.5° - 45° - 22.5° alternation of focusing (F), defocusing (D) and focusing (F) gradients is illustrated in **Figure 1**. Contrary to the design proposed in [8], which aims at achieving an achromatic bending, the gradient is small, below $g_{max}$= 10 T/m at top energy, corresponding to a normalized quadrupole strength $k_{max}$ of about 1.5 m$^{-2}$.

The geometry of the coils and the windings are such that the strength (either focusing or defocusing) per unit length is constant. Two power supplies are used: one for the quadrupole and one for the dipole layers.

The magnetic field was computed at the nominal current at a reference radius of 23 mm. In a curved system, an expansion in toroidal harmonics might be considered more appropriate but can be difficult to connect to beam optics requirements [11]. The use of local cylindrical harmonics followed by direct optimization of the coil has been shown to be sufficient for similar systems in the past [8].

The baseline design foresees a beam chamber at warm temperature, the coils inside a cryostat and the iron shielding again at room temperature.

The iron shielding around the coil is dimensioned such that – with an external diameter of the iron equal to 28 cm – the magnetic field in the iron does not exceed 1.6 T and the stray field is less 10 Gauss at 10 cm from the iron. These results have been obtained with a finite-element analysis in which the coil geometry was approximated with the ones of classical cosine-theta coils for the external dipole layer and of cosine-2-theta coils for the inner quadrupole layers. The advantage of this approximation, valid for shielding calculation, is that it does not require dense meshes for 3D studies of the coil performed with CST-studio software [11].

## 2.2 Gantry scanning system, structure and optics.

### Layout and scanning system

An important choice in the gantry design is the *downstream* position of the two scanning magnets (SMx and SMy) that minimizes the aperture of the last bending magnet(s). The drawback is that the space between the end of the last magnet and the gantry axis must be large enough to have more than 200 cm between the centre of the last SM (SMy) and the isocentre, i.e a Source to Axis Distance = SAD > 200 cm. This value is larger than the one of some commercial proton therapy gantries [13].

This constraint, together with the one of a total gantry outer radius not larger than 5.0 m, dictates the position of the three $90^0$ CCT magnets, the length of the drift spaces and the bending radius of $\rho$ = 1.65 m, corresponding to a bending field of 4.0 T. The gantry layout is shown in Figure 2.

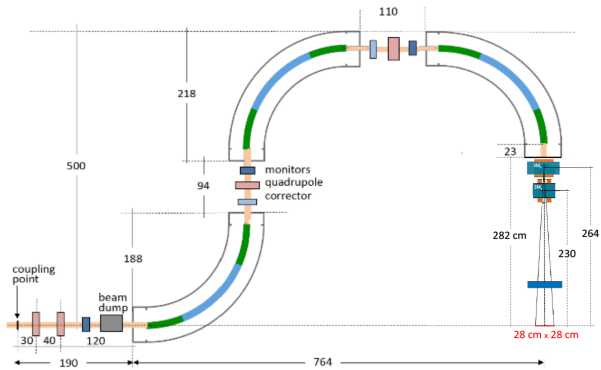

*Figure 2: Layout of the 4.0 T gantry: the quadrupole coils inside the main CCT magnets are highlighted in blue and green. External quadrupoles are in pink, correctors in ligth blue and instrumentation in dark blue.*

### Gantry mechanics

The gantry is designed to be compact, light and simple, therefore all structures that are not functional to the beam transport have been removed.

First of all, thanks to an electric motor with a high-torque planetary gear, which moves precisely the gantry to any angular position [14], the gantry does not need a counterweight. This choice reduces the weight by almost a factor two. Another advantage is that the momentum of inertia is reduced and thus the extra-stresses in case of deceleration e.g. due to an emergency stop. A further important simplification, with another consequent weight reduction, is that the magnets for the beam transport are structural elements. These two measures eliminate the need of the cradle, to support and guide the movement of the rotating part, and make it possible to attach the gantry to the wall. The rotation of $\pm 110^0$ is considered enough for most of the treatment plans [15] and for all of them if the patient bed is rotated by $180^0$. The attachment of the gantry to the wall leaves the floor ground level clean, which simplifies maintenance and repairs because a mobile platform (or a cherry picker device) gives access to any point of the gantry in any of its angular positions.

Finite-element studies with Autodesk Inventor Pro software [16] were performed to ensure that the mechanical stresses are well below the elastic limit and that the maximum displacement of the beam line is below 1 mm, when the gantry is horizontal and subject to the maximum torque.

The gantry rotating part has been conceived such that all its parts, including the magnets. can be assembled at the manufacturer premises. Thanks to its compactness, it is possible to transport it via a truck of appropriate capacity.

### Gantry beam optics

The optics code MADX [17] has been used to design, model and match the lattice of the gantry and of the synchrotron ring. The main $90^0$ AG-CCT magnet unit is modelled as four combined-function 'SBEND' magnets of $22.5^0$ each, carrying a gradient. Two short straight sections are also added at the beginning and end of each bending sequence to take into account the space needed for the return coils and the cryogenic vessel.

The goal of the optics design is to ensure that (i) the FWHM beam sizes at the isocentre range from 3 to 10 mm and (ii) the beam envelope fits within a maximum half-aperture of $A_x = A_y$ =20 mm, which is the radius of the superconducting magnets beam pipe, and in particular that:

$$\sqrt{\beta_{x,y}(s)\varepsilon_{Tx,y} + D_{x,y}^2(s)\delta_T^2} < A_{x,y} \qquad (1)$$

Where $\beta_{x,y}(s)$ is the beta function and $D_{x,y}(s)$ is the dispersion (zero in the non-bending plane). $\varepsilon_{Tx,y}$ are the total geometrical emittance, calculated as *five times* the root-mean-square (RMS) emittance ($\varepsilon_{Tx,y}$= 5 $\varepsilon_{rms,x,y}$) and $\delta_T$=2$\sqrt{5}$ $\delta_{rms}$ is the total relative momentum spread. Table 2 shows the expected emittance values in the extraction lines for the minimum and maximum extracted energies.

*Table 2: Beam parameters (total emittance and total momentum spread) assumed in the extraction line at the critical energies, from the PIMMS design* [18] [19].

|  | 100 MeV/u | 430 MeV/u |
|---|---|---|
| Horizontal emittance | 5 μm | 5 μm |
| Vertical emittance | 7.1 μm | 3.3 μm |
| Momentum spread | 1 10$^{-3}$ | 1 10$^{-3}$ |

It should be noted that, although the vertical emittance follows the expected adiabatic damping scaling, the horizontal emittance for the extracted beam is constant. This is indeed the case for a bar-of-charge slow-extracted beam [18] [19] from a synchrotron.

The gantry optics is matched to satisfy the set of constrains for point-to point imaging [18] [20] on the gantry transfer matrix $R_{ij}$ [17]. Provided that the entry beam is "round", i.e. it has the same horizontal and vertical size $\sqrt{\varepsilon_x \beta_x} = \sqrt{\varepsilon_y \beta_y}$, this set of constrains ensures that [20]: (i) the optics is achromatic ($R_{16} = R_{26} = 0$), (ii) the beam size is independent of the gantry rotation angle ($R_{12} = R_{34} = 0$), (iii) the magnification factor for the horizontal and vertical beam size is the same ($R_{11} = \pm R_{33} = \pm G$) with $G=1$, i.e. the beam is transported 1:1 from the gantry entrance to the isocentre. Since there is no constraint on $\alpha_x$ and $\alpha_y$, these are free parameters and can be varied to optimize the dimensions of the beam and the beam divergence at the isocentre.

## 2.3 Synchrotron layout and beam optics design

The compact synchrotron consists of four 90$^0$ AG-CCT magnets, similar to the gantry but of larger aperture, and it has a two-fold symmetry as shown in Figure 3.

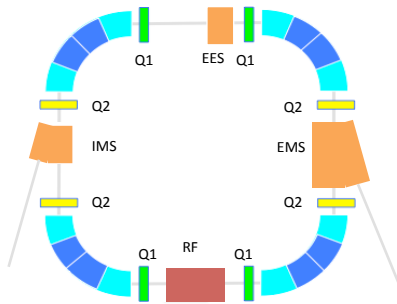

*Figure 3: Sketch of the synchrotron layout. AG-CCT magnets in blue, quadrupole families Q1,Q2 in green and yellow. Injection (IMS) and extraction septa (EES,EMS) in orange and RF cavities in red.*

To add tunability, two extra families of small fast, air-cooled, quadrupoles are added in the straight sections. The working point during the slow-extraction process is positioned close to a third order resonance, in this case the line $3Q_x = 5$. During injection and acceleration, the working point is safely far from it and it is optimized for an efficient multi-turn injection and to minimize space-charge effects. The accelerator is working below transition, as it is common for small rings.

Parametric studies have been done to identify the optimum value and length of the focusing and defocusing gradients of the AG-CCT, the length of the straight sections, and the strength of the air-cooled external quadrupoles. The goal was to minimize the beam size in the magnet round aperture, i.e. maximum beta-functions and dispersion, while staying below transition. Once a good parameter region was found with this method, the usual MADX matching routines are employed to find the optimum for each working point.

## 3 Results

### 3.1 Superconducting magnets

The AG-CCT magnets coils were designed to meet the specifications of Table 1, for the gantry and for the synchrotron. The two inner layers of alternating gradient quadrupole (AG-CCT) and the outer set of two CCT dipole layers have the radial distribution of Table 3.

*Table 3: Radial distribution of magnet layers.*

| Layer | Type | # wires /layer | Inner Radius (mm) | Outer Radius (mm) |
|---|---|---|---|---|
| Gantry | | | | |
| 1-2 | AG-CCT | 1 | 35.0 | 46.6 |
| 3-4 | CCT-dip | 9 | 46.6 | 66.8 |
| Synchrotron | | | | |
| 1-2 | AG-CCT | 2 | 55.0 | 70.2 |
| 3-4 | CCT-dip | 8 | 70.2 | 107 |

Each layer consists of a toroidal mandrel of cylindrical cross section with channels for the conductor. A radial width of 4 mm beneath each conductor channel is assumed for the mechanical support and layer-to-layer assembly gaps. A monolithic NbTi wire was chosen as the baseline conductor. The wire dimensions are 1.7 x 1.7 mm$^2$ (1.6 x 1.6 mm$^2$ bare); the copper to superconductor ratio is 2.8:1. Further optimization can explore different wire size based on desired operating current, magnet inductance, and mandrel fabrication. **Error! Reference source not found.** shows the load line of the gantry magnet: the number of wires per layer is set to guarantee at least 25-30% operating current margin.

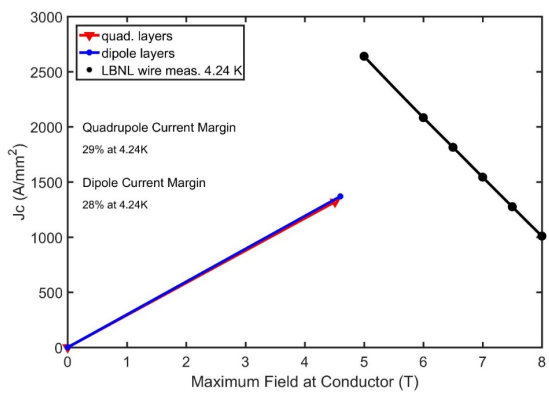

*Figure 4: Load line for the gantry magnet design*

The physical bending length corresponds to a $102^0$ angle for the gantry magnet and $106^0$ for the synchrotron magnet. At nominal current, the stored energy of both synchrotron and gantry magnet is about 0.4 MJ and the inductance slightly less than 1 H.

Because of the stringent requirements of the gantry weight and size, the iron shielding of the magnets has been optimized. The final outer radius of the iron shielding is 28 cm, and a cut of 8 cm has been applied on the top and bottom part of the magnet, as shown in Figure 5. The total weight of the iron is of only 3.7 tons, to which another 0.3 tons has be added for the coils.

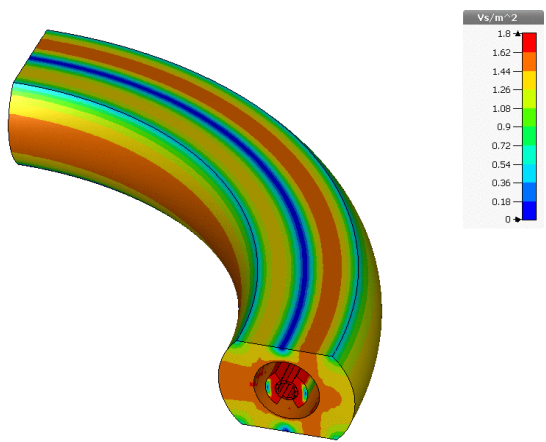

*Figure 5: 3D simulation for the optimized iron shielding. Magnetic field generated from the dipole and quadrupole coils, at their maximum strength as of Table 1, i.e. to generate 4 T and a gradient of 10 T/m in the vacuum chamber.*

The 3D magnetic simulations of Figure 5 show the field asymmetry expected from the strong curvature and confirm that the field inside the iron is smaller than 1.6 T, everywhere but in a small volume that requires further refinement.

### 3.2 Gantry structure, scanning system and optics

The main parameters of the gantry mechanics are summarized in Table 4.

*Table 4: Gantry mechanical design main parameters*

| Gantry **external** radius | **5**00 cm |
|---|---|
| Gantry length | 864 cm |
| Weight of each magnetic unit | 4 tons |
| Gantry weight | 35 tons |
| Center of mass distance | 243 cm |
| Max torque | 840 kNm |
| Gantry support weight | 15 tons |

A sketch of the gantry mechanical structure is shown in Figure 6, in which the main parts are identified.

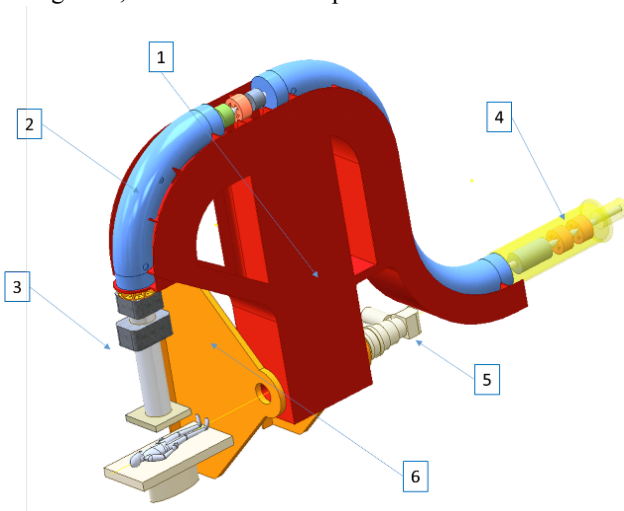

***Figure 6***: *Sketch of the gantry structure and its main parts: 1) rotating structure, 2) 90 degrees AG-CCT magnets, 3) delivery system: two Scanning Magnets, nozzle and monitors, 4) initial part of the vacuum chamber with quadrupoles and a beam dump, 5) motor and actuator that rotate the gantry by ±110° with respect to the horizontal plane, 6) part of the support, which is attached to the wall.*

The external radius of the gantry is 5 m and the weight of the rotating part of the gantry is 35 tons, mainly due to the 15 tons for the rotating support structure and about 4 tons for each of the three bending units.

This makes this *carbon ion* gantry smaller than the superconducting one installed at NIRS [2] (6.7 m radius), as compact as many *proton gantries* and even significantly lighter [6].

Industrial electric motors with planetary gears such as those of [14] can provide the required torque (840 kNm) and guarantee the required precision. Finite-element analysis assured that the deformations on the gantry are everywhere smaller than 1 mm. The maximum stresses are 28.5 MPa, very far from the 200 MPa limiting value. This means that the design is very safe as far as the nonlinear deformations are concerned, although the effect of soldering still needs to be evaluated.

Since the gantry is rotating, the cooling of the SC magnets will be provided by cryocoolers, with a limit on the ramp-rate. Two cryocoolers per magnet, with a capacity of 1.5 W (@4.2 K each [21] are considered and a small ramp-rate has been chosen as a "safe" margin, to minimize the dynamic heat load component:

$$\Delta B/\Delta t \leqq 0.15 \text{ T/s}. \quad (2)$$

For comparison, the superconducting gantry at NIRS [22] is equipped with 10 cryocoolers, 4 of them for their large-aperture $90^0$ magnet which has a maximum ramp-rate of 0.3 T/s.

Having the scanning system downstream of the last magnet implies challenging constrains in the design of the two SMs, because of the small SADs. A sketch of the system is shown in Figure 7 and the main parameters are collected in Table 5.

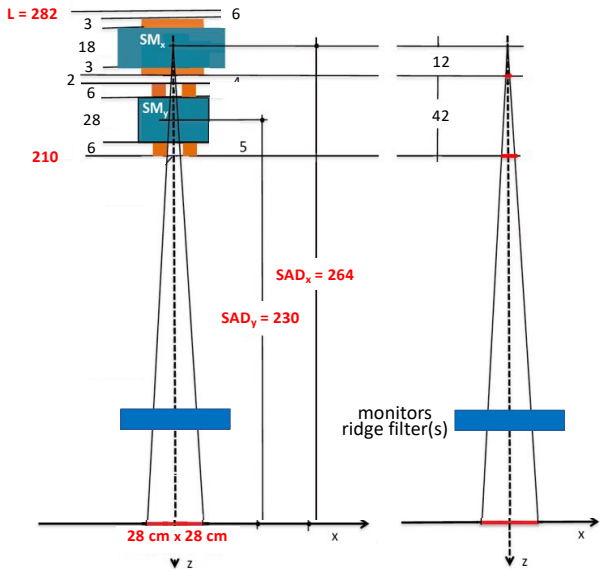

*Figure 7: Layout of the scanning system.*

*Table 5: Scanning Magnets parameters*

| Quantity | Unit | SMx | SMy |
|---|---|---|---|
| Source Axis Distance SAD | cm | 264 | 230 |
| Total length (with coils) | mm | 240 | 400 |
| Inter-pole gap | cm | 45 | 100 |
| Magnetic length | mm | 220 | 370 |
| Number of coils | | 2 | 2 |
| Number of turns/coils | | 30 | 60 |
| Peak current | A | 1000 | 750 |
| Inductance | mH | 3.4 | 17 |
| Resistance | mΩ | 45 | 115 |
| Peak field | T | ± 1.60 | ± 1.09 |
| Maximum bending angles | mrad | ± 53 | ± 61 |

The Table shows that the peak fields are 1.60 T for SMx and 1.07 T for SMy.

To scan the (28 cm x 28 cm) treatment area of the figure at an average speed of $v_{spot}$ = 20 mm/ms, the $SM_x$ ($SM_y$) currents $I_x$ ($I_y$) have to go from -1000A to +1000 A (-750 to +750) in 14 ms, so that the maximum rates are:

$$(\Delta I_x/\Delta t) = (\Delta I_y/\Delta t) \leqq 145 \ (110) \text{ A/ms}. \quad (3)$$

The scanning magnets of Figure 7 and Table 5 require fast and powerful power supplies and a very accurate control system. They can be built by properly modifying systems that have been developed at CERN for the Large Hadron Collider.

### Optics

Figure 8 shows the gantry Twiss functions, assuming initial conditions $\beta_x = \beta_y = 5$ m, $\alpha_x = \alpha_y = 0.8$.

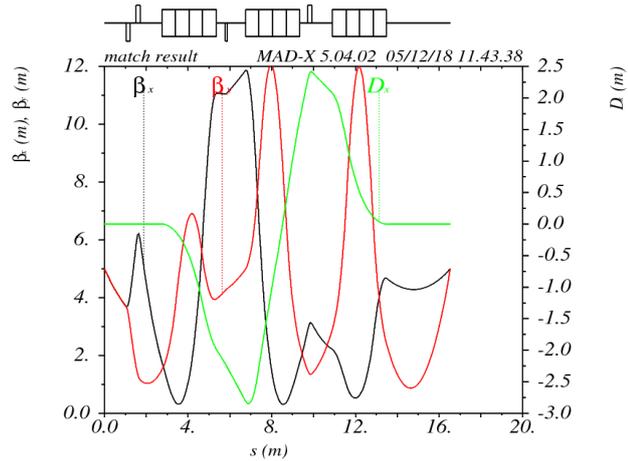

*Figure 8: Gantry optics. Initial $\beta_{x,y}$= 5 m, $\alpha_{x,y}$ = 0.8.*

The gradients of the AG-CCT are set independently for the three main magnets and, in addition, four extra quadrupoles guarantee the necessary degrees of freedom to fulfil the matching conditions. In a single pass system, such as a gantry, the beam size along the line depends on the initial beta functions, which are adjusted according to the required beam size at the isocentre.

Figure 9 shows the FWHM beam size, at 430 MeV/u, for the two extreme gantry rotation angles.

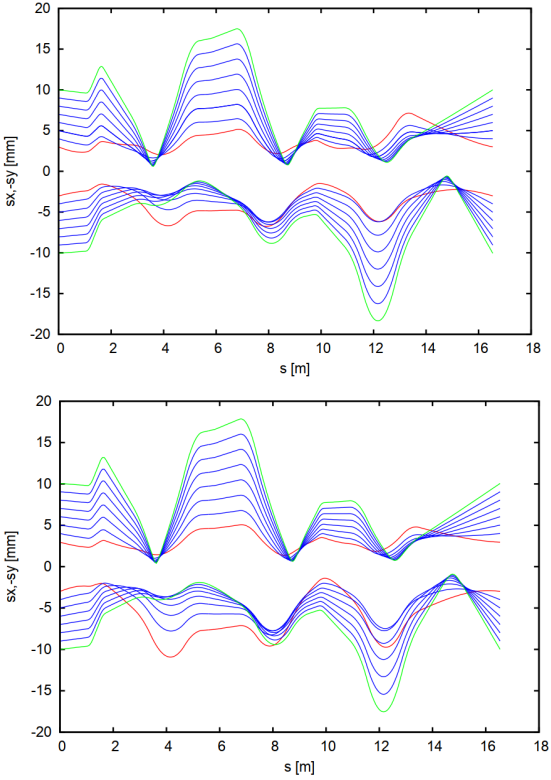

*Figure 9:* Horizontal and vertical beam size along the gantry for different beam spots at the isocentre, at 430 MeV/u. Top: angle = $0^0$, i.e. gantry horizontal. Bottom: angle = $90^0$, i.e. gantry vertical.

The plots are made for the different beam sizes (3-10 mm) at the isocentre. The beam has to be "round" at the gantry entry point, as discussed in the Section 2, and is transported 1:1 to the isocentre. The figures show that the beam transverse dimensions are everywhere smaller than the 20 mm radius of the vacuum chamber. Note that (i) all currents scale linearly with the beam momentum and (ii) the optics has not to be changed when either the beam size is adjusted in the range 3-10 mm or the gantry angle is changed between -110° and +110°.

### 3.3 Synchrotron layout and optics

The design of the synchrotron is profiting of the size reduction with respect to the conventional facilities, due to the use of SC magnets. A circumference of about 27 m has been obtained, which is about 3 times smaller than the CNAO and MedAustron accelerators.

Table 6 summarizes the main dimensions of the acceleration layout sketched in Figure 3 and the specifications for the external air-cooled magnets. The four $90^0$ CCT magnets have the specifications listed in Table 1.

*Table 6*: Accelerator dimensions and parameters.

| Circumference | 27 m |
|---|---|
| Injection energy | 7 MeV/u |
| Extraction energy | 100 → 430 MeV/u |
| Length of straight section 1 | 3 m |
| Length of straight section 2 le | 3.6 m |
| External quads length | 0.1 m |
| External quads max. strength | 10 T/m |

Figure 10 shows the lattice functions for the extraction optics.

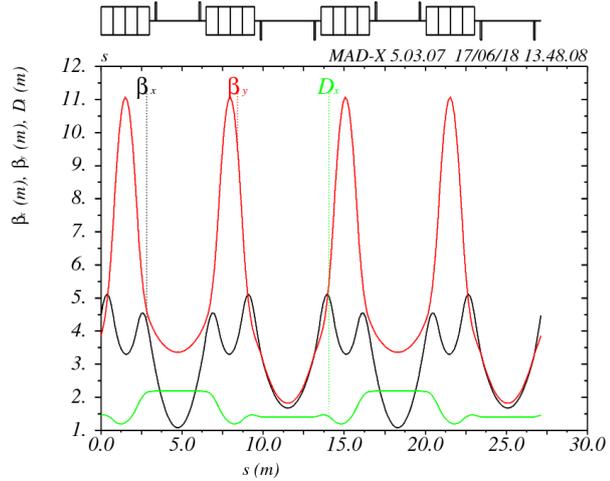

*Figure 10:* Optics functions at extraction (working point of $Q_x$ = 1.68, $Q_y$ = 1.13).

The horizontal tune is $Q_x$ = 1.68, to perform a resonant slow extraction on the third order line $3Q_x$ = 5, and can be moved far off the resonance, i.e. around $Q_x$ = 1.72 or further up, for injection and acceleration. Moreover, for extraction, the dispersion has been reduced as much as possible in the one of the straight sections 2, where the extraction septum is located.

Considering the maximum values for the optics functions, i.e. $\beta_x$ = 5 m, $\beta_y$ = 12 m and $D_x$ = 2.5 m and the beam properties summarized in Table 2, it follows that the maximum beam sizes are never larger than 13 mm and 18 mm, at injection. A magnet aperture of 60 mm diameter is sufficient to accommodate the beam, with margin for orbit excursions due to e.g. the injection bump.

## 4 Discussion

### 4.1 Facility footprint

Figure 11 shows a possible layout of a Carbon therapy facility [23], which includes the 27 m circumference synchrotron and two 5 m radius gantries. Figure 12 shows the comparison with existing facilities built on the PIMMS design [24] and the large footprint reduction obtained by using superconducting magnets.

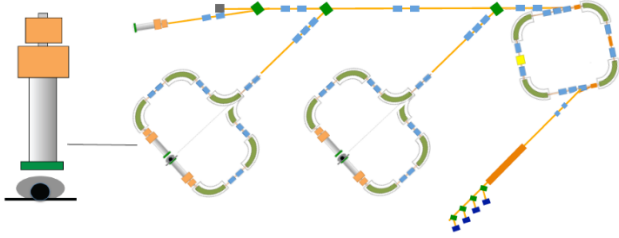

*Figure 11.* Schematic of the proposed facility layout [25] *(the shielding elements are not shown).*

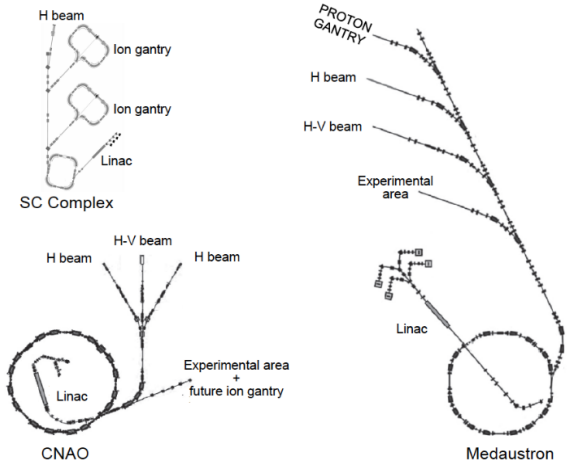

*Figure 12:* Comparison of the footprints of the proposed SC ion therapy facility with the one of CNAO and MedAustron.

The 7 MeV/u injector linac, which is about 10-m long, is fed by four ion sources. Indeed, in the future patients will be treated with different ions according to the depth and the radio-resistance of the tumour target.

### 4.2 Synchrotron magnets ramp-rate and multi-energy beam delivery

The time to ramp the SC synchrotron to the maximum beam energy is longer than 4 s, even for a magnet ramp rate of 0.8 T/s. This is to be compared with the typical time of 0.5 s of a NC-magnets synchrotron. To stay competitive, the full dose must therefore be delivered in a single cycle, with a scheme similar to the one implemented at HIMAC [26] and schematically shown by the black curve of Figure 13, which shows that the momentum of the extracted beam is varied in steps and the beam delivered in many energy flat-tops to 20-40 contiguous target layers that are typically 2 mm thick. The time to go from one flat-top to the next is typically in the range 100-200 ms.

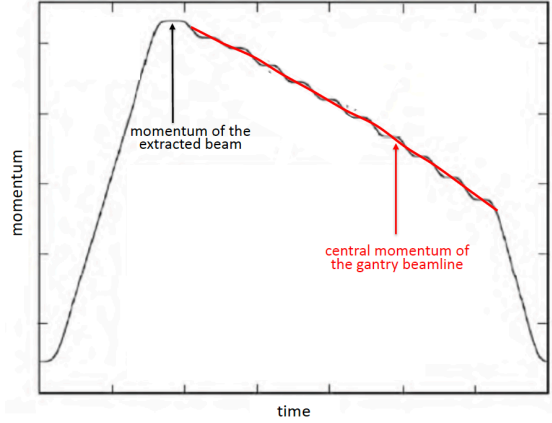

*Figure 13:* Multiple-energy operation cycle at HIMAC [26] *(black) and average slope of the ramp down (red) that could be the momentum of the gantry central line, as discussed in Section 4.4.*

### 4.3 Number of stored particles and injection issues

In order to treat a half-a-litre water equivalent tumour, $10^{10}$ Carbon-ions need to be stored and accelerated [24] in the SC synchrotron. Assuming that the injection lasts 50 turns with efficiency 50%, $2 \cdot 10^{10}$ C-ions need to be provided by the source in a pulse length of 37 μs, since the revolution period at 7 MeV/u injection energy is about 0.74 μs.

In the baseline design, an EBIS source produces high intensity, short pulse $C^{+6}$ ions. The production of $C^{+6}$ ions directly form the source has the advantage that the injector can be compact enough for acceleration up to 7 MeV/u. The facility can use the same RFQ and following HI structure which are under development for a medical Carbon ion linac [27]. The linac may eventually accelerate other ions with a ratio $Z/A = 1/2$, i.e. Helium ions, which are of particular interest of the medical community [28] and protons if powered at reduced voltage.

The EBIS source MEDeGUN is the compact source of $C^{+6}$ under development at CERN. It has a Brillouin-flow electron gun type and it is optimized to deliver short pulses of $10^8$ ions at high repetition rate (~200 Hz) for a medical hadron linac [29]. For a synchrotron, the requirements are different, i.e. a relatively long pulse of 37 μs with a low repetition repetition rate, but with an intensity higher by more than a factor 100. A possibility [30], which requires substantial R&D, is to keep the same EBIS electron current of 1 A from MEDeGUN and have a longer trap length (1 m instead of the 0.25 m) in order to produce a larger number of ions. Accumulating $10^{10}$ ions should be feasible, however at the limits. In order to increase the current, one may choose to have a larger trap, with 2A electron current, at the expenses of a larger emittance.

In case the modified, low-repetition-rate EBIS will not be able to produce the required ion current, ECR sources will be used. Even in this case, R&D is expected to increase the current by at lease a factor 2, as promised by the recent development e.g. of [31].

The third alternative for injection is an ECR source producing $C^{+4}$ ions, followed by acceleration and charge-exchange injection in the synchrotron. The method allows the accumulation of large currents in a smaller beam size, however it is not flexible if other ion species need to be injected. One could think, though, of designing a system with chicane magnets and a moving foil, which is able to inject both carbon and helium ions.

### 4.4 Gantry magnets ramp-rate and momentum acceptance

The ramp-rate of the gantry magnets can be very slow if the facility, equipped with a NC-magnet synchrotron, adopts the scheme of separate cycles for different extraction energies. In this case the time for the gantry magnets to reach the next energy level is about one cycle period, i.e. of the order of 1 s.

Assuming that the beam is delivered in multiple energy flat-tops according to the scheme of *Figure 13* (black line), a gantry magnet ramp-rate of less than 0.15 T/s - according to Eq. (2) - is still large enough to move from one energy level to the next in less than 200 ms. Indeed, in the worst case of a 100 MeV/u beam ($B = 2$ T) the water depth is $R = 70$ mm so that a 2 mm layer thickness corresponds to a fractional range variation $\Delta R/R = 3$ %. It follows $\Delta p/p \cong 1$ % and $\Delta B \cong 0.02$ T, according to the approximate formula

$$\frac{\Delta R}{R} \cong 3 \frac{\Delta p}{p} = 3 \frac{\Delta B}{B} \quad (3)$$

The red curve of *Figure 13* describes another possible method of delivering the dose: while the momentum of the extracted beam follows the black line, the currents in the gantry SC magnets are varied continuously and smoothly so that the central momentum accepted by the beam line follows the red curve. This is possible as long as the system accepts the small momentum offset corresponding to half layer, i.e. $\Delta p/p = \pm 0.5\%$. To this end the gantry is equipped with orbit correctors (positioned as shown in Figure 2) so that the off-momentum beam trajectory can be steered with the 3 correctors and then further adjusted with the horizontal scanning magnet SMx. These elements have - for a momentum offset of 0.5% - to provide maximum (extra) kicks of up to 5 mrad.

In a third more sophisticated approach to dose delivery - the Oblique Raster Scanning method proposed by one of the authors in [24] - the momentum of the beam extracted from the SC synchrotron changes continuously following the red line of Figure 13, as the central line of the gantry magnets.

Finally, as last consideration, having a gantry with a momentum acceptance of ±1% means that it is possible to perform a multiple-energy extraction of more than 5 energy steps spaced by 2 mm without changing the gantry magnet current. This is enough for the beam intensity available in most of the present medical facility.

### Conclusions

An innovative gantry structure has been described that keeps the weight of the rotating part below 35 tons. The major requirement to achieve such a light system is to have small-aperture superconducting magnets. It has been demonstrated that it is possible, with Alternating Gradient - Canted Cosine Theta (AG-CCT) magnets, to design an optics, invariant with respect to gantry rotation and achromatic, that fulfils the requirement of fitting within an aperture of maximum 20 mm radius, for beam sizes at the isocentre in the range 3 mm to 10 mm. A magnet ramp rate of up to 0.15 T/s is large enough to perform multi-energy extraction and beam trajectory correctors can be introduced to have a momentum acceptance equal to ±0.5%, which makes it possible for the SC magnets to vary continuously and not to follow strictly the beam momentum variation.

The AG-CCT magnet technology has been chosen for this gantry because it allows having nested alternating quadrupoles inside the main dipoles, independently powered, and therefore reduces the beam size and magnet dimensions.

A compact synchrotron has also been designed, using similar $90^0$ AG-CCT magnets with a larger aperture. The preliminary layout and optics have been presented and the injection system (including linac and sources), extraction and stored number of particles discussed in some detail.

### Acknowledgements

Discussions with A. Garonna and V. Ljubicic (TERA Foundation), G. Arduini, C. Carli, V. Parma, A. Pikin, J. Pitters and F. Wenander (CERN) have been fundamental in the preparation of this article.